\documentclass[aps,prl,floatfix,twocolumn,groupedaddress,showpacs]{revtex4}

\usepackage{hyperref}
\usepackage{amsmath}
\usepackage{epsfig, graphicx}

\begin{document}

\title{Suppression of Heating Rates in Cryogenic Surface-Electrode Ion Traps}

\author{Jaroslaw Labaziewicz}
\email[]{labaziew@mit.edu}
\author{Yufei Ge}
\author{Paul Antohi}
\author{David Leibrandt}
\author{Kenneth R. Brown}
\author{Isaac L. Chuang}
\affiliation{Massachusetts Institute of Technology, Center for Ultracold Atoms, Department of Physics, 77 Massachusetts Avenue, Cambridge, MA, 02139, USA}

\date{\today}

\begin{abstract}
Dense arrays of trapped ions provide one way of scaling up ion trap quantum information processing.
However, miniaturization of ion traps is currently limited by sharply increasing motional state decoherence at sub-$100~\mu\text{m}$ ion-electrode distances.
We characterize heating rates in cryogenically cooled surface-electrode traps, with characteristic sizes in $75~\mu\text{m}$ to $150~\mu\text{m}$ range.
Upon cooling to $6~\text{K}$, the measured rates are suppressed by $7$ orders of magnitude, two orders of magnitude below previously published data of similarly sized traps operated at room temperature.
The observed noise depends strongly on fabrication process, which suggests further improvements are possible.
\end{abstract}

%\keywords{}

\maketitle

%\section{\label{sec:Intro}Introduction}
Quantum information processing offers a tantalizing possibility of a significant speedup in execution of certain algorithms\cite{Shor:97, Grover:97a}, as well as enabling previously unmanageable simulations of large quantum systems\cite{Feynman:82, Porras:04a}.
One of the most promising avenues towards practical quantum computation uses trapped ions as qubits.
Interaction between qubits can be mediated by superconductive wires\cite{Tian:04}, photons\cite{Duan:04, Moehring:07} or by shared phonon modes\cite{CZ:95}.
The last scheme has been most successful so far, having demonstrated one and two qubit gates\cite{Monroe:95a}, teleportation\cite{Riebe:04,Barrett:04}, error correction\cite{Chiaverini:04a} and shuttling\cite{Hensinger:06}.
Scaling of these experiments to a large number of ions will require arrays of small traps, on the order of $10~\mu\text{m}$, to achieve dense qubit packing, improve the gate speed and reduce the time necessary to shuttle ions between different traps in the array\cite{Wineland:98, Kielpinski:02, Steane:07, Slusher:05}.
Micro-fabrication techniques have been successfully used to fabricate a new generation of ion traps, demonstrating trap sizes down to $30~\mu\text{m}$\cite{Monroe:06, Seidelin:06, Britton:06}.
However, as the trap size is decreased, ion heating and decoherence of the motional quantum state increases rapidly, approximately as the fourth power of the trap size\cite{Turchette:00, Deslauriers:04, Deslauriers:06}.
% Using usual scaling: 
% Current sec freq ~1MHz*(100um/trapsize)
% Heating rate is ~(2000*(1MHz/trapfreq)^2)*(100um/trapsize)^4 = ~2000*(100um/trapsize)^2
% heating rate/sec freq = .002*(100um/trapsize)
At currently observed values, the heating rate in a $10~\mu\text{m}$ trap would exceed $10^6~\text{quanta/s}$, precluding ground-state cooling or qubit operations mediated by the motional state.

The strong distance dependence of the heating rate suggests that the electric field noise is generated by surface charge fluctuations, which are small compared to the distance to the ion.
Charge noise is also observed in condensed matter systems, where device fabrication has proven critical in reducing the problem\cite{Martinis:04, Oh:06}.
Similar advances in ion traps are impeded by lack of data and models accurately predicting measured noise\cite{Lamoreaux:97, James:98b, Henkel:99a}.
The charge fluctuations have been demonstrated to be thermally driven, providing another plausible route to reduce the heating.
Cooling of the electrodes to $150~\text{K}$ has been shown to significantly decrease the heating rate\cite{Deslauriers:06}.

In this Letter, we present the first measurements of heating rates in ion traps cooled to $6~\text{K}$.
We designed and built a range of surface-electrode traps, in which we are able to cool a single ion to motional ground state with high fidelity and observe heating on a quantum level.
Although the traps have very high heating rates at room temperature, these rates are suppressed by $7$ orders of magnitude at $6~\text{K}$, to values $2$ orders of magnitude lower than in previously reported traps of similar size operated at room temperature\cite{Turchette:00, Deslauriers:04, Deslauriers:06, Seidelin:06}.
Surprisingly, the heating rate exhibits an extremely strong dependence on the thermal processing of the trap.
These results strongly indicate that improvements in fabrication will allow for further suppression of the heating.

\begin{figure}\includegraphics[width=3.4in]{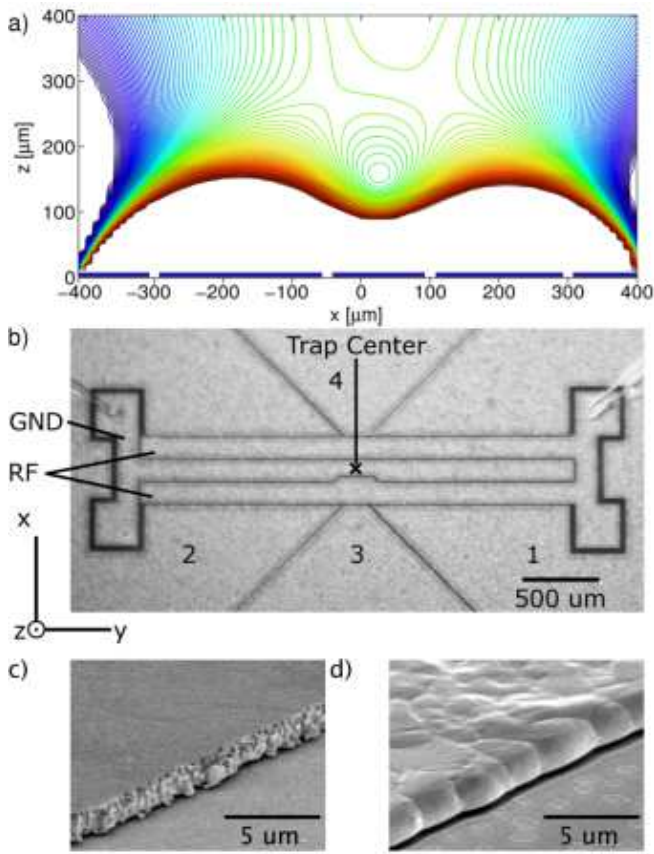}
\caption{\label{Fig:Trap}(color online) (a) Computed pseudo-potential in the x-z plane at the trap center for $\text{V}_1 = \text{V}_2 = 16~\text{V}$, $\text{V}_{3} = -16~\text{V}$, $\text{V}_{4} = -13~\text{V}$, $\text{V}_{RF} = 240~\text{V}_{amp}$. Isosurfaces are separated by 50mV.
The trap electrodes are outlined above x axis.
(b) Microscope image of the electrode geometry for the $150~\mu\text{m}$ trap.
DC electrodes are numbered from 1 to 4.
The notch in the central electrode tilts the principal axes of the trap by $6^\circ$ to $20^\circ$, depending on DC voltages, and defines a point where the RF field is zero.
In this trap, the width of the central electrode at the notch is $136~\mu\text{m}$, with electrode spacing of $21~\mu\text{m}$.
The potential minimum is $150~\mu\text{m}$ above the surface.
(c) SEM image of one of the trap electrodes before annealing and (d) after annealing at $760~^\circ\text{C}$ for 1 hour.}
\end{figure}

%\section{Planar Trap Design and Fabrication}
The traps are of a five rod surface-electrode design\cite{Chiaverini:05}, which facilitates thermal anchoring and scalability to multi-zone traps with complex geometries.
The simulated pseudo-potential above the surface and the electrode geometry are shown in Fig.~\ref{Fig:Trap}.
We fabricate the traps on a single crystal quartz substrate, a high thermal conductivity material at cryogenic temperatures, with a single step etch.
First, we evaporate $10~\text{nm}$ of Ti sticking layer, followed by $1~\mu\text{m}$ of Ag, chosen for its favorable etch properties.
The wafer is coated with NR9--3000 photoresist and patterned using optical lithography.
Exposed areas are etched away using $\text{NH}_3\text{OH}:\text{H}_2\text{O}_2$ silver etch, followed by HF Ti etch.
The etching process results in rough edges with multiple sharp points which emit electrons via Fowler-Nordheim tunneling when driven with voltages in excess of $100~\text{V}$.
The emitted electrons induce stray electric fields, eventually opening the trapping potential.
In order to smooth the electrode edges, we anneal the traps in vacuum ($10^{-5}~\text{torr}$), in a copper oven heated by a tungsten filament, at $720~^\circ\text{C}$ to $760~^\circ\text{C}$ for 1 hour.
The high temperatures used are necessary to re-flow the silver, but result in an optically roughened surface.
% Usual RF drive is 150mV resulting in 250V on trap. I check for glow up to 500mV, but it flattens some.
After annealing, we find no field emission points for voltages up to $750~\text{V}$ across the gaps.
Typical inter-electrode spacing is $\approx 10~\mu\text{m}$.
Finished traps are glued to a ceramic pin grid array carrier, and mounted on the $4~\text{K}$ plate of a bath cryostat filled with liquid helium. 
%\cite{Antohi:07}.
Good thermal contact is provided using a strip of copper mesh soldered to the trap surface and connected to the helium bath.
$1~\text{nF}$ filter capacitors placed close to the electrodes reduce RF pickup and noise.
Two-stage RC filters with cutoffs at $4~\text{kHz}$ cooled to $4~\text{K}$ remove noise from DC sources.
We fabricated a range of traps with designed electrode sizes at the center of $75~\mu\text{m}$, $100~\mu\text{m}$ and $150~\mu\text{m}$.
Due to a rather large inter-electrode spacing, we did not fabricate traps smaller than $75~\mu\text{m}$.
The numerically calculated trap height above the surface is within $3\%$ of the electrode size.

%\section{Sideband Cooling and Heating}

The traps are typically operated at a RF voltage of $250~\text{V}$ and frequency of $26~\text{MHz}$ ($150~\mu\text{m}$ trap) or $38~\text{MHz}$ ($100$ and $75~\mu\text{m}$ traps).
Typical DC electrode voltages satisfy $\text{V}_1 = \text{V}_2 = -\text{V}_3 = -\frac{5}{4}\text{V}_4$, where $\text{V}_i$ is the voltage on electrode $i$ (Fig.~\ref{Fig:Trap}).
% Check
For $\text{V}_1 = 25~\text{V}$ in the  $150~\mu\text{m}$ trap, the secular frequencies are $1~\text{MHz}$ along y and $2.5~\text{MHz}$ and $2.3~\text{MHz}$ in the x-z plane.
At these settings, the heat dissipated by the RF drive increases the trap temperature to about $6~\text{K}$, as measured using a $\text{RhO}$ temperature sensor glued to the chip.

Single $^{88}\text{Sr}^+$ ions are loaded from an ablation plume produced by a Q-switched, frequency tripled Nd:YAG laser incident on a $\text{SrTiO}_3$ target\cite{Kwong:89, Hashimoto:06}.
The ion is Doppler cooled on the dipole-allowed $S_{1/2} \leftrightarrow P_{1/2}$, $422~\text{nm}$ transition to $< 1~\text{mK}$, followed by sideband cooling of the lowest frequency mode on the quadrupole-allowed $S_{1/2} \leftrightarrow D_{5/2}$, $674~\text{nm}$ transition\cite{Leibfried:03, Letchumanan:06}.
We use pulsed sideband cooling, where the ion is illuminated with a pulse of $674~\text{nm}$ light on the red sideband of the $S_{1/2},~m=-\frac{1}{2} \leftrightarrow D_{5/2},~m=-\frac{5}{2}$ transition, followed by re-pumping to the ground state via the $P_{3/2}$ state using a $1~\mu\text{s}$ pulse of $1033~\text{nm}$ light.
This cycle is repeated $\approx 150$ times with progressively longer $674~\text{nm}$ pulses, starting at $10~\mu\text{s}$ and going up to $25~\mu\text{s}$, cooling the ion to motional ground state with fidelity exceeding $95\%$.
We interleave cooling pulses with pulses on the $S_{1/2},~m=\frac{1}{2} \leftrightarrow D_{5/2},~m=-\frac{3}{2}$ transition, to optically pump the ion to the $m=-\frac{1}{2}$ state.
The laser system used in this experiment is based on optical feedback to low finesse transfer cavities and described in Ref.~\cite{Labaziewicz:07}.
Long term drifts are removed by locking the lasers addressing the $S_{1/2} \leftrightarrow P_{1/2}$ and the $S_{1/2} \leftrightarrow D_{5/2}$ transitions to the trapped ion.

\begin{figure}\includegraphics[width=3.4in]{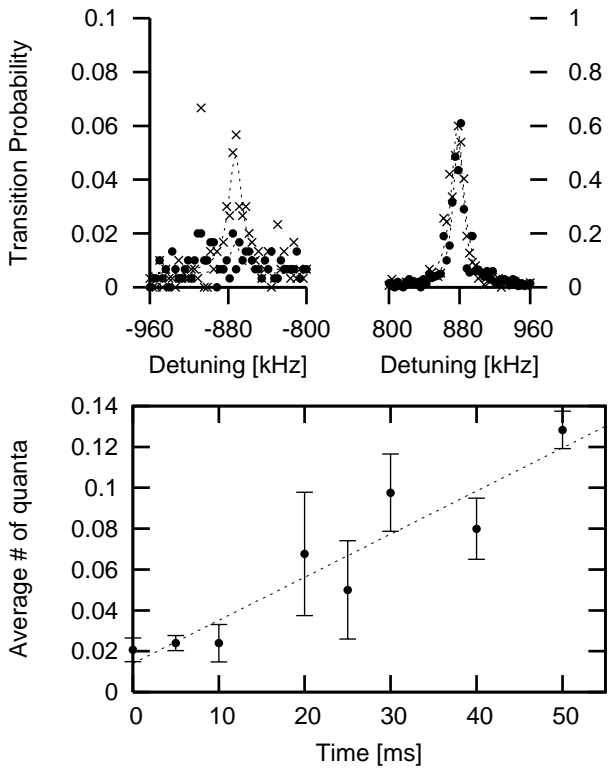}
\caption{\label{Fig:HeatingTime} Sample data taken in the $150~\mu\text{m}$ trap at $6~\text{K}$.
(top) Sideband spectra after cooling (solid circles) and after a $40$ ms delay (crosses).
Note that the scale for red (left) sideband is $10$ times smaller than the scale for blue (right) sideband.
(bottom) Average number of quanta versus delay time, with a linear fit.
The slope of the fit line is $2.1\pm0.2~\text{quanta/s}$ }
\end{figure}

Temperature measurement is accomplished by probing both blue and red sidebands of the $S_{1/2},~m=-\frac{1}{2} \leftrightarrow D_{5/2},~m=-\frac{5}{2}$ transition.
% using a $\approx65~\mu{s}$ long pulse, equivalent to $13~\pi$ on the carrier (NEEDS WORK).%
The state of the ion is subsequently detected by measuring the light scattered on the $S_{1/2} \leftrightarrow P_{1/2}$ transition with a PMT.
The sidebands are broadened due to laser linewidth and instability, and therefore fit to a Gaussian.
The average number of quanta is estimated using the ratio of the sideband heights\cite{Turchette:00}.
In order to determine the heating rate, we delay the readout, and measure the average number of quanta versus the delay time.
Typical data are shown in Fig.~\ref{Fig:HeatingTime}.

%\section{Heating Rate Data and Discussion}

Heating rates depend on the ion mass and its secular frequency in the trap.
To remove this dependency, we compute the electric field noise density $S_{E}(\omega)$ at the secular frequency of the ion using
\[
S_{E}(\omega) = \frac{4 m \hbar \omega}{q^2} \dot{n}
\]
where $m$ is the ion mass, $q$ is the ion charge, $\omega$ is the motional frequency and $\dot{n}$ denotes the actual heating rate in quanta per second\cite{Turchette:00}.
Furthermore, we normalize the field noise to $1~\text{MHz}$, assuming that the noise has $\omega^{-1}$ power dependence\cite{Deslauriers:06}.

\begin{figure}\includegraphics[width=3.4in]{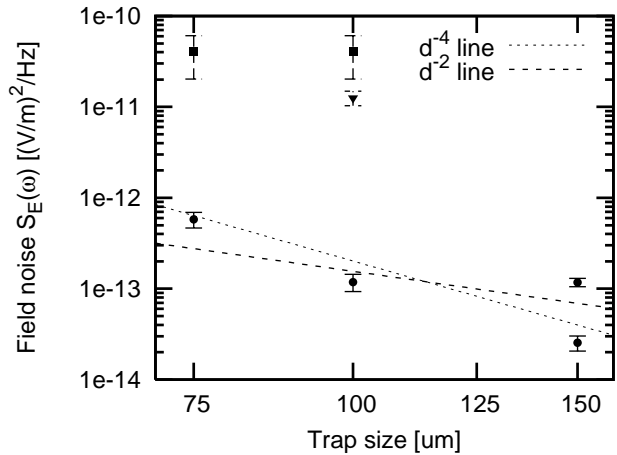}
\caption{\label{Fig:NoiseTraps} Measured noise density in $7$ different traps, with secular frequencies in $0.85 - 1.25~\text{MHz}$ range.
Each data point is normalized to $1$ MHz, assuming the spectrum scales as $\omega^{-1}$.
Circles correspond to traps annealed at $760~^\circ\text{C}$, squares correspond to traps annealed at $720~^\circ\text{C}$ while the triangle corresponds to a trap fabricated on sapphire.}
\end{figure}

Fig.~\ref{Fig:NoiseTraps} shows the measured field noise density in 7 different traps.
Closed circles correspond to traps annealed at $760~^\circ\text{C}$, squares correspond to traps annealed at $720~^\circ\text{C}$, while the triangle corresponds to a trap fabricated on a sapphire substrate, using the same process, and annealed at $760~^\circ\text{C}$.
The silver electrodes' properties depend strongly on the annealing temperature.
In silver on quartz traps, at $760~^\circ\text{C}$ we observed re-crystallization and void formation in the Ag film.
To characterize film roughness, we measured the fraction of power of a $650~\text{nm}$ laser reflected from the surface at normal incidence.
The reflectivity of the annealed film was reduced from initial value of $>90 \%$ to $\approx 5 \%$.
At $720~^\circ\text{C}$, the Ag film was much smoother, with reflectivity $\approx75 \%$.
At temperatures exceeding $760~^\circ\text{C}$ most of the silver film evaporates during the annealing process.
No other difference in fabrication or processing could be discerned; the Ag film did not undergo any chemical reactions observable using x-ray photon spectroscopy.
Due to different annealing properties of silver on sapphire, when annealed at $760~^\circ\text{C}$ the surface of this trap had similar properties to the silver on quartz traps annealed at $720~^\circ\text{C}$.

There are two significant trends in the data.
% Summary of heating rates I quote
% Turchette:00
% S = 1e-11 (Mo, 175), S = 4e-9 (Be, 175um), S = 3.4e-9 (Au, 280um)
% Deslauriers:04
% points around S = 1e-11 (200um)
% Deslauriers:06
% S = 7e-12 (Cd, 150um)
% Seidlin:06
% S = 1.6e-10 (40um) or S = 8.3e-13 at 150
% Our trap
% S = 2.4e-14 (150um)
The scaled electric field noise measured in quartz traps annealed at $760~^\circ\text{C}$ is more than $2$ orders of magnitude lower at $6~\text{K}$ than most reported room temperature traps and a factor of $30$ better than the best value\cite{Turchette:00, Deslauriers:04, Deslauriers:06, Seidelin:06}.
%The results are repeatable, as shown by the similar heating rates in two distinct $150~\mu\text{m}$ traps.
There is a clear increase of the heating in smaller traps, but due to systematic errors from the use of distinct traps and low number of points, the data is consistent with both $d^{-2}$ and $d^{-4}$ scaling.
Surprisingly, we find that traps with highly reflective surfaces have field noise $2$ to $3$ orders of magnitude higher than similar traps annealed at $760~^\circ\text{C}$.
The trap fabricated on sapphire shows similarly high heating rates, implying that the insulating layer is not the dominant source of noise.

In order to draw a better comparison with room temperature systems, we fabricated two separate $150~\mu\text{m}$ silver on quartz traps annealed at $760~^\circ\text{C}$.
The field noise at room temperature was measured in three separate ways, using Doppler re-cooling\cite{Seidelin:06}, boil-off time from the trap, and minimum laser intensity sufficient to cool the ion.
% Room temperature noise level
% S = 4e-7 (150um) at room temp
Using these methods we estimate the normalized field noise to be $30\pm7\times10^{-8}$ and $25\pm6\times10^{-8}~\text{V}^2/\text{m}^2/\text{Hz}$, $7$ orders of magnitude higher than the value we measured at $6~\text{K}$.
Given the very wide range of observed heating rates in other experiments, our values, though $\approx 50$ times higher than some of the previously reported values\cite{Turchette:00}, are not inconsistent with those experiments.
The base pressure in the system was $5\times10^{-10}~\text{torr}$, and is not expected to contribute to heating appreciably.

\begin{figure}
\includegraphics[width=3.4in]{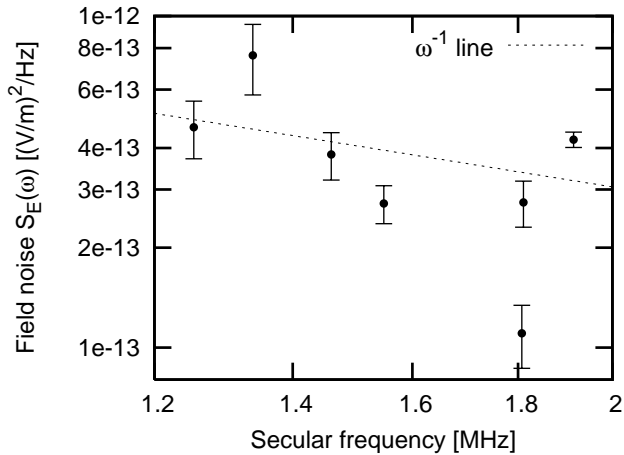}
\caption{\label{Fig:NoiseFrequency} Measured field noise density in a $75~\mu\text{m}$ trap at different frequencies of the ion motion.
The dependence is consistent with $\omega^{-1}$ scaling.}
\end{figure}

Fig.~\ref{Fig:NoiseFrequency} shows the measured electrical noise density as a function of trap secular frequency.
We were able to investigate only a narrow range of frequencies due to reduced stability of the trap outside of that range.
However, in that range, we find that $S_{E}(\omega)$ scales as $\omega^{-1}$, and the heating rate $\dot{\bar{n}}$ scales as $\omega^{-2}$.
This result is consistent with the literature, and justifies the $\omega^{-1}$ scaling used in Fig.~\ref{Fig:NoiseTraps}.

The heating rates in traps annealed at the higher temperature are very low, but still exceed the expected values.
At frequencies near $1~\text{MHz}$, the DC filters used have output impedance around $10~\Omega$.
We estimate the resistance of the leads from DC filters to the trap electrodes to be $<1~\Omega$, with the trap electrodes having a negligible resistance.
Assuming the noise is dominated by the Johnson noise of the filters at $4~\text{K}$, the expected voltage noise on the DC electrodes is estimated to be $S_V(\omega) = 2\times10^{-21}~\text{V}^2/\text{Hz}$.
The RF electrode has a low resistance connection to ground at ion secular frequencies.
Moreover, when properly compensated, the electric field due to voltage on the RF electrode is zero at the ion position, and therefore we do not expect it to contribute significantly to the heating.
% Geometric factor:
% Field is 200V/m for 1V on the endcap in 100um trap
% Effective endcap distance is 5mm
The electric field noise at the ion is estimated, using boundary element modeling of the electrostatic fields, to be $S_E(\omega) = S_V(\omega)\left(200\frac{100\mu\text{m}}{d}\right)^2 = 1\times10^{-16}\left(\frac{100\mu\text{m}}{d}\right)^2~\text{V}^2/\text{m}^2/\text{Hz}$, where $d$ is the trap size at the center.
The excess noise is most likely due to microscopic patch potentials\cite{Turchette:00, Deslauriers:06}.

Extrapolating our data to $10~\mu\text{m}$ traps operating at $10~\text{MHz}$ secular frequency, we expect heating rates of $\approx1000~\text{quanta/s}$, low enough to perform high fidelity operations\cite{Monroe:95a}.
The demonstration of surface electrode traps operating at $6~\text{K}$ opens up the possibility of integrating ion traps and superconductive systems\cite{Tian:04}.
Our results strongly suggest that an improved fabrication process will allow for further suppression of the noise in surface electrode ion traps.
The sensitivity of ions to charge fluctuations makes such experiments relevant to understanding of charge noise in other charge-based systems.

\begin{acknowledgments}
This work was supported by the Japan Science and Technology Agency and the NSF Center for Ultracold Atoms.
\end{acknowledgments}


\end{document}